\documentstyle[aps,psfig,eqsecnum,twocolumn]{revtex}
\def\di{\displaystyle}

\def\step{}

\def\eq#1{(\ref{#1})}
\def\Eq#1{Eq.~(\ref{#1})}

\def\Tr{{\rm Tr}}

\def\s0#1#2{\mbox{\small{$ \frac{#1}{#2} $}}}
\def\0#1#2{\frac{#1}{#2}}

\def\eq#1{(\ref{#1})}
\def\Eq#1{Eq.~(\ref{#1})}

\makeatletter

\renewenvironment{thebibliography}[1]
         {\frenchspacing\small
          \begin{list}{[\arabic{enumi}]}
         {\usecounter{enumi}\parsep=2pt\topsep 0pt
         \settowidth{\labelwidth}{[#1]}
         \leftmargin=\labelwidth\advance\leftmargin\labelsep
         \rightmargin=0pt\itemsep=0pt\sloppy}}{\end{list}}

\makeatother

\begin{document}
\twocolumn[\hsize\textwidth\columnwidth\hsize\csname
@twocolumnfalse\endcsname
\title{Perturbation theory and renormalisation group equations}

\author{
\hfill 
Daniel F. Litim\,${}^*$
and 
Jan M. Pawlowski\,${}^\dagger$
\hfill
\raisebox{23mm}[0mm][0mm]{\makebox[0mm][r]{
CERN-TH-2001-327\ \ FAU-TP3-01-10\ \ {\tt hep-th/0111191}}}%
}
\address{${}^*${\it 
Theory Division, 
CERN, 
CH-1211 Geneva 23.
}\\ ${}^\dagger${\it 
Institut f\"ur Theoretische Physik III, 
Universit\"at Erlangen, 
D-91054 Erlangen.
}}

\maketitle

\begin{abstract}\noindent
  {We discuss the perturbative expansion of several one loop improved
    renormalisation group equations. It is shown that in general the
    integrated renormalisation group flows fail to reproduce
    perturbation theory beyond one loop. 
    \\

{PACS: 05.10.Cc,\ 11.10.Gh\ \ \
E-mail: Daniel.Litim@cern.ch\,, \ jmp@theorie3.physik.uni-erlangen.de}

}
\end{abstract}
\vskip2.pc]

\noindent 
{\bf 1. Introduction}\\[-1ex]

\noindent
Renormalisation group (RG) methods are an essential ingredient in the
study of non-perturbative problems in continuum and lattice
formulations of quantum field theory.  A number of RG equations have
been proposed, where the starting point is the (infrared) regularised
one loop effective action. Taking the derivative w.r.t.\ the infrared
scale together with a subsequent one loop improvement leads to a flow
for the effective action. The merit of such an equation is its
flexibility, as it allows for non-perturbative approximations not
bound to the weak coupling regime. Thus, these flows are particularly
interesting for theories where one has to resort to truncations
because the full problem is too hard to attack. Indeed, surprisingly
good results concerning critical exponents in scalar theories have
been obtained within simple approximations to a particular version of
a one loop improved RG \cite{Litim:2001hk}, based on a proper-time
representation of the one loop effective action \cite{Liao:1996fp}. It
has also been suggested that the proper-time RG may be an interesting
tool for gauge theories, since the regularisation respects a local
non-Abelian gauge symmetry \cite{Liao:1997nm}.\step

However, results obtained within a truncated system are only as good
as the accompanying quality checks. Apart from the inherent problems
of these checks, the present situation requires additional care, since
most of the one loop improved RG lack a first principle derivation.
Such flows suffer from a severe conceptual problem.  It is unclear,
whether they are only approximations to flows for the full effective
action or whether they represent an exact flow. The latter is indeed
known to hold true for Exact RG (ERG) flows \cite{Polchinski,CW}
(for reviews see \cite{Bagnuls:2000ae}). They can be obtained within a
one loop improvement, but also from a first principle derivation,
mostly done within a path integral representation.  The strength of
exact RG flows is that systematic approximations of the integrated
flow correspond to systematic approximations to the full quantum
theory. This property, in combination with the convergence behaviour
of the flow, is at the root of the predictive power of the
formalism. The similarity of the different one loop improved flows,
including ERG flows, has fuelled hopes that the scenario just
described for exact flows may be valid in general.  \step

Based on this picture, and prior to an application of a general
one loop improved flow to any physical problem, it is mandatory to
either prove that a given flow is exact, or to unravel its inherent
approximations. A way to settle these questions consist in a detailed
comparison of one loop improved flows with known exact flows.  Within
the derivative expansion, this has been studied in
\cite{Litim:2001hk}.  In this note, we take a different route and
study one loop improved RG equations within perturbation theory.  It
is shown that they only represent, in general, approximations to flows
in the full theory.  This result is achieved by a structural analysis
of the flows, and by calculating the diagrammatic representation of
the two loop contributions to the effective action generated by the
flow through an iterative formal integration.  In general neither the
graphs nor the combinatorial factors of the two loop diagrams that
originate from one loop improved flows, are the correct ones. A full
account of the present calculation together with a discussion of
related issues will be presented in \cite{consistent}.
\\[2ex]

\noindent 
{\bf 2. One loop improved renormalisation group}\\[-1ex]

\noindent
We briefly review the philosophy of a one loop improved
renormalisation group. The starting point is the formal equation for
the one loop effective action:
\begin{eqnarray}\label{1loop}
\Gamma^{\rm 1-loop}=S_{\rm cl}
+\s012 \Tr\ln S^{(2)}\,.
\end{eqnarray}
The trace in \eq{1loop} is ill-defined and requires -at least- an UV
regularisation. A one loop improved RG is derived from \eq{1loop} by
first employing an explicit regularisation, taking the derivative
w.r.t.\ the cut-off scale $k$ and then substituting $S^{(2)}$ by
$\Gamma^{(2)}$. Here, we
concentrate on infrared regularisations; this does not make a
difference for the flow itself, which in either case
 should be local in momentum space, e.g.~only a small momentum
range about $q^2\approx k^2$ contributes to the flow at fixed $k$.
\step

Let us start with the derivation of the ERG flow
\cite{Polchinski,CW,Bagnuls:2000ae}.  Adding an infrared regulator $R$
(a momentum dependent mass term) to $S^{(2)}$ in \eq{1loop} and
proceeding according to the one loop improvement philosophy, we arrive
at
\begin{eqnarray}\label{exact}
\partial_t \Gamma_k=\s012 \Tr \left(\Gamma_k^{(2)}+R\right)^{-1}
\partial_t R\, ,
\end{eqnarray} 
where $t=\ln k$ is the logarithmic infrared scale introduced via $R$.
The regulator $R$ has to meet some requirements as a function of
momentum and the cut-off scale, which are discussed at length in the
literature. For our purpose these consistency requirements are
irrelevant, since we only want to perform iterative formal
integrations. 
\step

We emphasise that a general exact flow is the flow of some operator
insertion within the theory. A first principle derivation of the ERG,
for example, is based on the insertion $\s012 \int \phi R\phi$.
Insisting on the one loop nature of the flow, one is {\it bound} to an
insertion which is at most quadratic in the fields. Otherwise, the
corresponding exact flow would also contain higher loop contributions. 
We conclude that an exact flow with a
one loop structure must depend {\it linearly} on the full propagator.
This is indeed the case for the ERG flow \eq{exact}.\step

Another possibility for regularising the expression in \eq{1loop} is
to modify the trace itself by inserting an operator
$\rho$ multiplicatively \cite{Liao:2000yu}. This amounts to the
replacement $\Tr \ln S^{(2)}\to \Tr \rho \ln S^{(2)}$ in \eq{1loop}
and leads to the one loop improved RG flow
\begin{eqnarray}\label{operator}
\partial_t \Gamma_k=\s012 \Tr\ \partial_t\rho\ \ln \Gamma_k^{(2)}\,. 
\end{eqnarray} 
The multiplicative structure of this flow is particularly convenient,
when used in numerical applications. Note, that opposed to \eq{exact}, 
the flow \eq{operator} depends on the logarithm of $\Gamma_k^{(2)}$. 
Based on this structure, we can already conclude that \eq{operator}
cannot be exact. \step

Finally we consider a regularisation based on a proper-time
representation of \eq{1loop},
\begin{eqnarray}\label{RegDef}
\Gamma^{\rm 1-loop}=S_{\rm cl}
-\s012 \int\0{ds}{s} \ \Tr\ \exp \left(-s\,S_{\rm cl}^{(2)}\right)\,.
\end{eqnarray} 
Now we multiply the integrand in \eq{RegDef} by a regularising
function $f(s\,\Lambda^2)-f(s\,k^2)$ \cite{Oleszczuk:1994st}.
Proceeding along the lines of the one loop improvement we arrive at
\cite{Liao:1996fp}
\begin{eqnarray}\label{PTRG}
\partial_t \Gamma_k= \s012 \int_0^\infty \0{ds}{s}\
\partial_t f \ \Tr\exp\left(-s\Gamma^{(2)}_k\right)\,.
\end{eqnarray} 
In order to facilitate the perturbative calculations below, we cast
the flow equation \eq{PTRG} in a form which is more convenient for
this purpose. This alternative representation also reveals more clearly
the structure of the proper-time flows. To that end, we expand a
general proper-time flow in the following basis set of regulator
functions $f$:
\begin{eqnarray}
\label{fm}
\partial_t f(x;m)& = & \0{2}{\Gamma(m)} x^m  \exp\left({-x}\right)\,. 
\end{eqnarray}
Here, $x=k^2 s$. Note that the IR behaviour is controlled by the term
$e^{-x}$, where $x$ serves as a mass. These flows cover all
proper-time flows that have been studied in the literature
\cite{Litim:2001hk,Liao:1996fp,Liao:1997nm,Floreanini:1995aj,Schaefer:1999em,Meyer:2000bz,Papp:2000he,Bohr:2001gp,Meyer:2001zp,Bonanno:2001yp,Mazza:2001bp,Zappala:2001nv}.
Moreover, linear combinations $\sum_m d_m\, f(x;m)$ of \eq{fm} with 
$\sum_m d_m=1$ cover all flows with mass-like IR behaviour. 
The trace in \eq{PTRG} can be written in terms of the normalised
eigenfunctions $\Psi_n$ of $\Gamma^{(2)}_k$ with $ \Gamma^{(2)}_k
\Psi_n=\lambda_n \Psi_n$. Within this representation we deal with
simple $s$-integrals. By performing the $s$-integration we arrive at
\cite{consistent}
\begin{eqnarray}\label{PTRG-CS}
\partial_t \Gamma_k=\Tr \left({k^2\over \Gamma_k^{(2)}+k^2}\right)^m. 
\end{eqnarray}
The operator kernel inside the trace is the $m$th power of a
Callan-Symanzik kernel. We note that the {\it functional dependence}
of \eq{PTRG-CS} on $\Gamma^{(2)}$ depends on the regularisation.
Above, we have argued that an exact one loop flow has to depend
linearly on the full propagator. Hence, \eq{PTRG-CS} is not exact for
$m\neq 1$ due to the non-linear dependence of \eq{PTRG-CS} on the full
propagator.\step

In addition, \eq{PTRG-CS} also signals that, at least in perturbation
theory, the deviation of a general proper-time flow from an exact flow
is regularisation-dependent. In contrast, for both \eq{exact} and
\eq{operator}, the functional dependence on $\Gamma^{(2)}$ and,
thus, the result of their formal integration is independent of
the regularisation.  For \eq{PTRG}, however, linear combinations of
\eq{PTRG-CS} span the space of all kernels which decay at least as
$(\Gamma^{(2)}+k^2)^{-1}$ and reproduce the one loop effective action.
A general kernel trivially leads to a non-unique endpoint of
the flow. This result also implies that \eq{PTRG}, in general, is not
an exact flow.
\\[2ex]

\noindent 
{\bf 3. Effective action at one loop}\\[-1ex]

\noindent 
Thus, prior to any use of the flows \eq{operator} and \eq{PTRG}, it is
mandatory to collect more information on their inherit deviation from
exact flows. Here, this is done by explicitly calculating one loop and
two loop effective actions following from the flows. 
This also serves as an independent proof of our general statements. 
We restrict ourselves to a scalar
theory with one species of fields, but with general interaction.  The
results are easily generalised to arbitrary field content. As the
flows \eq{exact}, \eq{operator} and \eq{PTRG} are derived as one loop
improved flows from the one loop effective action \eq{1loop}, their
integrals reproduce the one loop effective action in the limit, where
the infrared cut-off tends to zero.  It is instructive to see how this
comes about.  The one loop contribution $\Delta\Gamma_1$ is given by
\begin{eqnarray}\label{formal1loop} 
\Delta\Gamma_1= \int_\Lambda^k {d k'\over k'} 
\left(\partial_{t'}\Gamma_{k'}\right)_{\rm 1-loop}\, .
\end{eqnarray}
Here, $\left(\partial_{t'}\Gamma_{k'}\right)_{\rm 1-loop}$ 
stands for the right-hand sides in either of the flow equations \eq{exact},
\eq{operator} or \eq{PTRG}, with $\Gamma_k^{(2)}$
substituted by $S^{(2)}$. This is sufficient to obtain the effective
action at one loop.  \step

Consequently, integrating the ERG flow \eq{exact} leads to 
\begin{eqnarray}\label{exact1loop} 
\Delta\Gamma_1= 
\s012 \Tr \, \left[\ln (S^{(2)}+R)\right]_\Lambda^k\, . 
\end{eqnarray}
Note that even for $k\neq 0$ the expression functionally resembles the
one loop contribution to the effective action. Indeed, it is the UV
regularised one loop contribution for a theory with propagator
$S^{(2)}+R$.\step

Integrating the one loop improved flow \eq{operator} leads to 
\begin{eqnarray}\label{operator1loop} 
\Delta\Gamma_1=
\s012\Tr \left[\rho\,\ln S^{(2)}\right]_\Lambda^k\,.  
\end{eqnarray}
Again this resembles the one loop effective action for any $k$. In
contrast to an ERG flow, however, it is impossible to
interpret \eq{operator1loop} as the one loop contribution of 
an UV-regularised modified theory.\step

Integrating the proper-time flow \eq{PTRG-CS} at one loop, we get 
after a straightforward algebra 
\begin{eqnarray}
\Delta\Gamma_1=\s0{1}{2 m}\Tr \left[ \left(\s0{k'^2}{
S^{(2)}}\right)^m 
{}_2 F_1\left(m,m;m+1;-
\s0{k'^2}{S^{(2)}}\right)\right]_\Lambda^k,\!\!\!  
\label{PTRG-1loop}\end{eqnarray}
where ${}_p F_q(x,y;z;w)$ is the generalised hyper-geometric series.
For integer $m$, the series in ${}_2 F_1$ in \eq{PTRG-1loop} can be
summed up and there is a simpler representation for the one loop
contribution:
\begin{eqnarray}
\Delta\Gamma_1=\s012\Tr \left[ \ln \left(\mbox{\small $S^{(2)}+{k'}^2$}\right)
-\sum_{n=1}^{m-1} \s01n 
\left(\s0{{k'}^2}{ S^{(2)}+{k'}^2}\right)^n\right]_\Lambda^k. 
\label{PTRG-m}\end{eqnarray}
For $k\neq 0$ \eq{PTRG-1loop} does not resemble the one loop
contribution to the effective action. Of course, for $k\to 0$,
\eq{PTRG-1loop} reproduces the one loop effective action
$\s012[\Tr\ln(S^{(2)}+{k}^2)]_{\rm ren}$ where the
renormalisation at $\Lambda$ is included. \\[2ex]

\noindent 
{\bf 4. Effective action at two loop}\\[-1ex]

\noindent
As the ERG flow \eq{exact} has a first principle derivation,
obviously it has to reproduce the correct two loop result.
Structurally it belongs to the same class as the usual Callan-Symanzik
flow, and the calculation of diagrams and combinatorial prefactors of
either flow goes along the same lines. Here, we only present the result
of such a calculation. The two loop contribution $\Delta\Gamma_2$ to
the effective action obeying \eq{exact} is given by
\begin{eqnarray}
\nonumber 
\lefteqn{\Delta\Gamma_2 = 
\int_{pp'qq'}
\Bigl[ {1\over 8}\,   G_{pp'}\ S_{p'pqq'}^{(4)}\ G_{q'q}} 
\hspace{1.4cm}\\
& &\di   -{1\over {12}}\,\int_{ll'} 
G_{pp'}\ S_{p'lq}^{(3)} \ G_{ll'}\ 
S_{l'pq'}^{(3)}\ G_{q'q}
\Bigr]_{\rm ren.}\, , 
\label{profield} 
\end{eqnarray}
where the subscript ${}_{\rm ren.}$ indicates that these are
renormalised diagrams due to the subtractions at $\Lambda$.  We have
introduced the abbreviations $G_{pp'}\equiv(S^{(2)}+R)^{-1}(p,p')$,
the vertices $S_{p_1\cdots p_n}^{(n)} \equiv \delta^{(n)}S /
\delta\phi(p_1) \cdots \delta\phi (p_n)$, and a convenient short-hand
notation for the momentum integrals $\int_{p_1\cdots p_n}\equiv\int
\s0{d^dp_1}{(2\pi)^d}\cdots \s0{d^dp_n}{(2\pi)^d}$.  The combinatorial
factors in \eq{profield} are in agreement with perturbation
theory. Again, even for $k\neq 0$ the result \eq{profield} functionally
resembles the perturbative structure. This analysis can be
easily extended to any loop order.  Note that one can always rewrite
the integrands as total $t'$-derivatives. Thus, the precise form of the
regulator $R$ is irrelevant for the result, as it should.\step

Expanding the one loop improved flow equation \eq{operator} at two
loop leads to the following expression:
\begin{eqnarray}\label{sb2loop} 
\Delta\Gamma_2=\s012 \int_\Lambda^k \0{dk'}{k'}
 \int_{pp'qq'}\ \Delta{\Gamma^{(2)}_{1,pp'}}\ G_{p'q}\ 
\partial_{t'} \rho_{qq'}
\end{eqnarray}
and $G=1/S^{(2)}$.  It is easy to rewrite the expression
on the right hand side of \eq{sb2loop} as a total derivative, since the
only $k$-dependence of $\Delta \Gamma_1^{(2)}$ is given by $\rho$. 
We finally get
\begin{eqnarray}\nonumber 
\lefteqn{\Delta\Gamma_2=  
\int_{pp'qq'}
\Bigl[ {1\over 8}\,   
(G\, \rho)_{pp'}\ S_{p'pqq'}^{(4)}\ (G\, \rho)_{q'q}}\hspace{.75cm}\\
& &\di       -{1\over 8}\,\int_{ll'} 
(G\, \rho)_{pp'}\ S_{p'lq}^{(3)} \ G_{ll'}\ 
S_{l'pq'}^{(3)}\ (G\, \rho)_{q'q}
\Bigr]_{\rm ren.}. 
\label{profieldsc} 
\end{eqnarray}
Again, as for \eq{profield}, the result does not depend on the
regulator for $k=0$, where $\rho= 1$. Differentiating
\eq{profieldsc} w.r.t.~$k$ leads to the integrand of \eq{sb2loop}, as
it should.  The combinatorial factors of the diagrams in
\eq{profieldsc} do not match those in \eq{profield}. Thus the flow
\eq{operator} fails to reproduce perturbation theory beyond one loop.
\step

Finally we discuss the proper-time flow \eq{PTRG}. Below \eq{PTRG-CS},
we have already argued that the flow \eq{PTRG} is not an exact flow
for a general regulator. Here, as an explicit example, we calculate
the two loop effective action for $m=2$.  Expanding the flow \eq{PTRG}
at two loop we get
\begin{eqnarray}\label{2-loop} 
\Delta\Gamma_2 =\! -2 \int_\Lambda^k \0{dk'}{k'} \int_{pp'}
\ \Delta\Gamma_{1,pp'}^{(2)}\ (G\ k'^2\ G\ k'^2\ G)_{p'p} , 
\end{eqnarray} 
where $G_{pp'}\equiv (S^{(2)}+k'^2)^{-1}(p,p')$. Note, that it is
impossible to rewrite the integrand in \eq{2-loop} as a total
derivative w.r.t.\ the scale parameter $t'$.  This already is a strong
hint at the fact that one cannot get the correct two loop result.  Let
us cast \eq{2-loop} in a form which shows explicitly how it deviates
from perturbation theory. Using partial $t'$-integrations we obtain
from \eq{2-loop}, after some lengthy but straightforward algebra,
\begin{eqnarray}\nonumber
\Delta\Gamma_2 &=&
\int_{pp'qq'} \Bigl[ \018\,   G_{pp'}\ S_{p'pqq'}^{(4)}\ G_{q'q}
\\ && \nonumber \di
\qquad\quad
-\0{1}{12}\int_{ll'} G_{pp'}\ S_{p'lq}^{(3)} \ G_{ll'}\ 
S_{l'pq'}^{(3)}\ G_{q'q}\Bigr]_{\rm ren.}
\\  \di \nonumber  
&-&
\012 \, \int_\Lambda^k \0{dk'}{k'} \int_{pp'qq'll'}
\Bigl[(G\ k'^2\, G)_{pp'} \ S_{p'ql}^{(3)}
\\ &&\di 
\qquad \qquad \times(G\ k'^2\, G)_{qq'} \ S_{q'pl'}^{(3)}\
(G\ k'^2\, G)_{l'l}\Bigr] \,. 
\label{deviation}
\end{eqnarray}
Differentiating \eq{deviation} w.r.t.~$k$ leads to the integrand of
\eq{2-loop}, as it should.  The first two terms in \eq{deviation}
correspond to the correct two loop result as presented in
\eq{profield}. The last term denotes the deviation from standard
perturbation theory. The $d\ln k'$-integrand of the last term in
\eq{deviation} is the non-standard diagram depicted in Fig.~1.  The
last term on the right-hand side of \eq{deviation} cannot be absorbed
in renormalisation constants. It contains arbitrary powers in fields
and momenta and does not integrate to zero in the limit $k\to0$ and
$\Lambda\to\infty$. For massive theories both limits are safe.
Consequently this term displays a non-trivial deviation of the
proper-time flow from perturbation theory. The form of the integrand
is that of the sunset graph where all propagators have been
substituted by their squares.  This is clearly related to the fact
that the form of the proper-time flow is that of a Callan-Symanzik
flow with all propagators substituted by their squares.

\begin{figure}
\begin{center}
\unitlength0.001\hsize
\begin{picture}(600,430)
\psfig{file=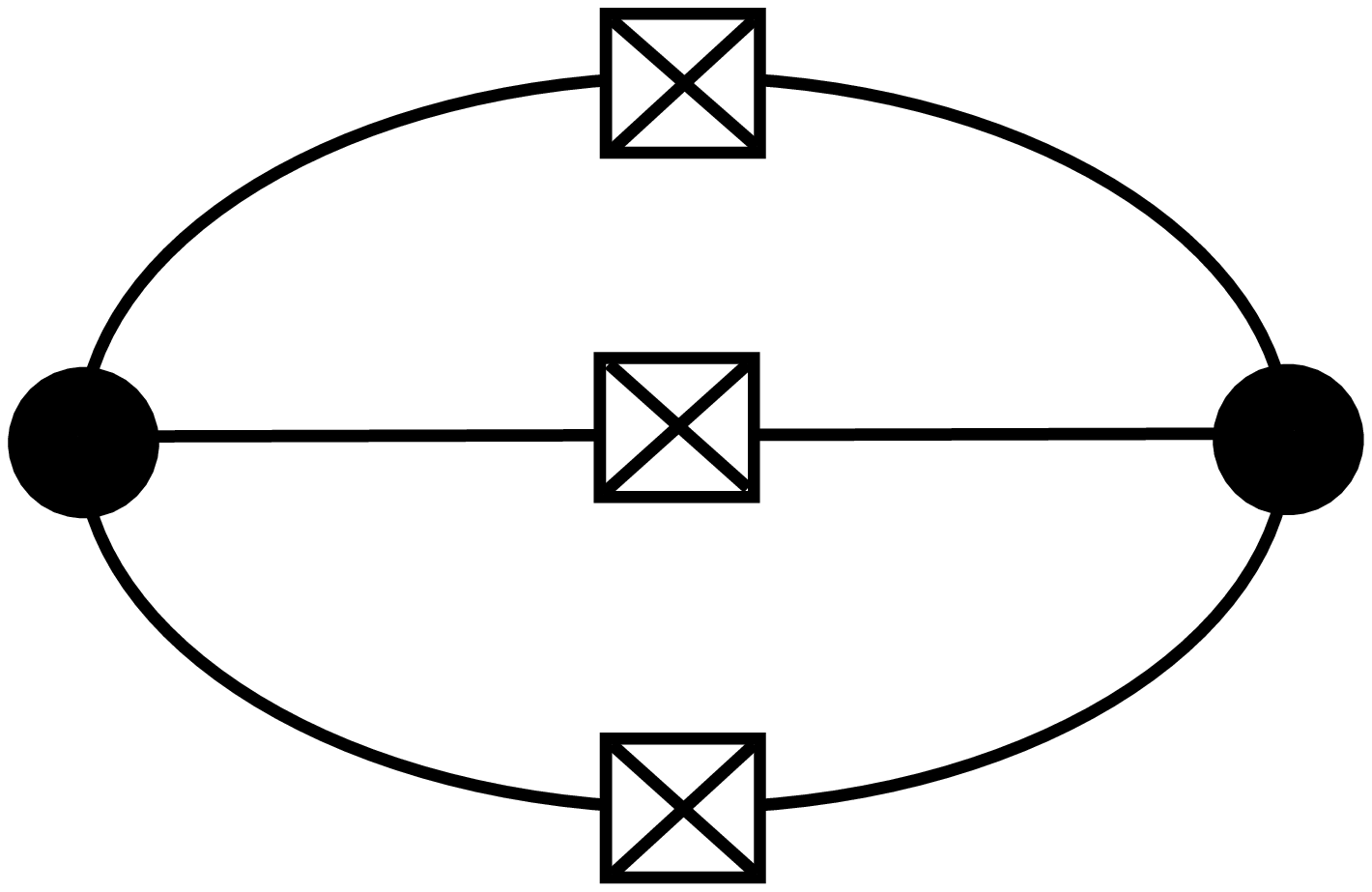,width=.55\hsize}
\end{picture}
\begin{minipage}{\hsize}{
{ \small {\bf Figure 1:} The integrand of the non-standard term in
  \eq{deviation}. The two vertices $S^{(3)}$ are denoted
  by $\bullet$, the six internal lines are the propagators
  $G=(S^{(2)}+k'^2)^{-1}$, and the three insertions correspond to
  $k'^2$.}}
\end{minipage}
\end{center}
\vskip-3ex
\end{figure}

To be more explicit, consider the example of a massive $\phi^4$-theory
with mass $M$ and quartic interaction $ \s0{\lambda}{4!}\, \int d^d x
\,\phi^4$. The contribution of the non-standard diagram to the
propagator is obtained after taking the second derivative with respect
to the fields in \eq{deviation} at $\phi=0$. We find
\begin{eqnarray}\nonumber 
\lefteqn{\lambda^2 \int_\infty^0 \0{d k}{ k} \int \frac{d^d q}{(2\pi)^d}
\frac{d^d l}{(2\pi)^d}\,\Bigl[  {k^2\over (k^2+M^2+q^2)^2}}
 \hspace{.6cm} 
\\ & &\di 
\times {k^2\over (k^2+M^2+l^2)^2} 
{k^2\over (k^2+M^2+(l+q-p)^2)^2}\Bigr]\, . 
\label{propphi4} \end{eqnarray}
The integrand it strictly positive. Hence the integral is
non-vanishing.  Moreover it has a non-trivial momentum dependence.
This can be seen by evaluating the limits $p\to 0$ and $p\to\infty$.
For $p\to 0$ we are left with a non-vanishing constant. In turn, for
$p\to\infty$ the expression in \eq{propphi4} vanishes.\\[2ex]

\noindent 
{\bf 5. Discussion}\\[-1ex]

\noindent
Having established that neither \eq{operator} nor, in general,
\eq{PTRG} provide exact flows, we want to understand what precisely
causes the deviation from perturbation theory. First we recall the
argument made prior to \eq{operator}: A general exact flow is related
to the flow of an operator insertion in the theory. Demanding,
additionally, that the flow has a one loop structure restricts
possible insertions to operators quadratic in the fields. Consequently
such a flow has to depend linearly on the full propagator. \step

For a general flow it might be hard to decide, whether one has such a
situation. Already for general proper-time flows we had to take the
detour of expanding general flows in the basis \eq{fm} in order to
reach to a conclusion. Thus, we would like to provide an additional
criterion, which also reflects the necessity of a linear dependence on
the full propagator.  Indeed, a sufficient condition for a RG equation
to reproduce perturbation theory can be deduced from the iterative
structure of the perturbation series: It suffices that the solution of
a RG equation has the same iterative structure even at non-vanishing
cut-off. Without this property, the corresponding RG equation has to
satisfy an infinite tower of iterative constraints in order to
reproduce perturbation theory in the limit, where the infrared cut-off
tends to zero.  Consequently, one can assess from the structure of the
one loop effective action at $k\neq 0$, whether a flow is likely to
reproduce perturbation theory.  \step

The iterative structure discussed above is absent in the one loop
effective action given in \eq{PTRG-1loop} for $k\neq 0$. Moreover, it
cannot be regained by considering linear combinations of regulators
\eq{fm}.  Despite this discouraging fact, let us shed some more light
on the structure of proper-time flows. It is not possible to integrate
a general proper-time flow beyond one loop without knowing the precise
form of the regulator.  Still, there are recursive relations between
different proper-time flows at a given loop order.  These relations
tell us how the flows differ from each other for {\it arbitrary} $m$,
integer or not.  At two loop, and with $G=(S^{(2)}+k^2)^{-1}$, the most
general recursion relation is given by \cite{consistent}
\begin{eqnarray}\nonumber 
\Delta\Gamma_{2,m}&-&\Delta\Gamma_{2,m-1}=
\012\int_\infty^0 
\0{dk}{k}\,\Tr \big[\left({G\,k^2}\right)^{m-1}\, G
\\ \di 
&& \times\, \displaystyle 
(\s0{m}{m-1}\,{k}^2 G-1)
\,{\delta^2\over(\delta\phi)^2}\Tr\, (G \,{k}^2)^{m-1}
\big]\,,
\label{recursive} \end{eqnarray}
apart from irrelevant terms from the different renormalisation
procedures for the two flows.  The difference \eq{recursive} (or, more
generally, $\Delta\Gamma_{2,m}-\Delta\Gamma_{2,m-n}$ with integer $n$)
depends on arbitrarily high powers of the fields and does not
integrate to zero. \step

\Eq{recursive} can be used to give an independent explicit proof of
the non-exactness of general proper-time flows. To that end, let us
assume for a moment that the proper-time flow for a particular $m_0$
is exact. Then it follows from \eq{recursive} that all flows with
$m=m_0+n$ for integer $n$ are {\it not} exact, because the
corresponding terms \eq{recursive} do not vanish identically in the
fields. Hence, of all proper-time flows of the form \eq{PTRG} with
regulators \eq{fm} or finite linear combination thereof, the set of
exact flows is of measure zero.  This has an immediate consequence for
flows with integer $m$. The Callan-Symanzik flow ($m=1$) is exact, but
any flow with integer $m>1$, or any linear combinations thereof, are
not exact.  Hence, the structure of the findings for $m=2$ is present
for arbitrary $m$, and \eq{recursive} provides an independent explicit
proof for the general statement derived after \eq{PTRG-CS}.

Thus, for proper-time flows, we arrive at the following picture. The
only known exact proper-time flow is the Callan-Symanzik flow. Other
exact proper-time flows -if they exist- would require a linear
dependence on the full propagator, possibly in some disguise. Based on
our findings, no further exact flows can be found within the set of
regulators \eq{fm}, which covers all flows previously studied in the
literature. Of course, it is not excluded, that a regulator, which is
represented by an infinite series of regulators \eq{fm}, is exact.
However, there is no {\it a priori} criterion upon which one could
embark on and construct such a regulator.  \step

To summarise, we have shown that the one loop improved flows
\eq{operator} and, in general, \eq{PTRG} are not exact flows. We have 
shown explicitly, that they fail at the first non-trivial order, 
at two loop. These results imply 
that hopes expressed in the literature -- suggesting that the
RG flows \eq{operator} and \eq{PTRG} correspond to exact flows only
with a different implementation of the regularisation -- cannot be
maintained.  In fact, these flows are {\it substantially} different
from exact flows, and describe at best approximations to the latter.
Justification of their use requires a deep understanding of the
inherent approximation in order to furnish these methods with
predictive power.  This question has only been addressed within the
derivative expansion \cite{Litim:2001hk}.  However, the potential
benefits of general one loop improved RG flows within numerical
implementations justify further investigations.  An extensive study of
this problem, including a more detailed account of the present
calculations, will be given elsewhere \cite{consistent}.
\\

{\it Acknowledgements:} JMP thanks CERN for hospitality.  DFL has been
supported by a Marie-Curie fellowship under EC contract
no.~HPMF-CT-1999-00404.
\\[2ex]
\newpage
\noindent 
{\bf References}\\[-1ex]


\end{document}